\documentclass[manuscript, nonacm, screen=true]{acmart}

\usepackage{longtable}
\usepackage{capt-of}
\usepackage{balance,bm}
\usepackage{color, colortbl, xcolor}
\definecolor{LightGray}{gray}{0.97}
\usepackage{pgfplots}
\usepackage{xcolor}
\usepackage{url}
\usepackage{makecell}
\usepackage{multirow}

\usepackage{subcaption}
\usepackage{booktabs} 
\usepackage{graphicx}
\usepackage{textcomp}
\usepackage{color,soul}
\usepackage{bm}
\usepackage{multirow}
\usepackage{wrapfig}
\usepackage{balance}
\usepackage{graphicx}  
\usepackage{enumitem}

\usepackage{colortbl}
\usepackage{arydshln}
\usepackage [english]{babel}
\usepackage [autostyle, english = american]{csquotes}
\definecolor{linkColor}{RGB}{6,125,233}
\definecolor{green}{rgb}{0.0, 0.65, 0.31}
\definecolor{bleudefrance}{rgb}{0.19, 0.55, 0.91}
\definecolor{ceruleanblue}{rgb}{0.16, 0.32, 0.75}
\definecolor{grey}{HTML}{969696}
\definecolor{violet}{HTML}{756bb1}
\definecolor{dgrey}{HTML}{01665e}
\definecolor{lgrey}{HTML}{5ab4ac}
\definecolor{dgreen}{HTML}{005a32}
\definecolor{purple}{HTML}{ae017e}

\definecolor{editCol}{HTML}{000000}
\definecolor{maskCol}{HTML}{c51b7d}
\definecolor{lrColor}{HTML}{8856a7}
\definecolor{trColor}{HTML}{d01c8b}
\definecolor{ctColor}{HTML}{4dac26}
\definecolor{brickred}{HTML}{f03b20}
\definecolor{improveCol}{HTML}{253494}
\definecolor{worsenCol}{HTML}{d7191c}
\definecolor{DarkBlue}{HTML}{00008B}
\definecolor{mscolor}{HTML}{01665e}
\definecolor{nmscolor}{HTML}{bf812d}
\definecolor{lgreen}{HTML}{ccece6}
\definecolor{dolive}{HTML}{308014}

\definecolor{lrColor}{HTML}{8856a7}
\definecolor{trColor}{HTML}{d01c8b}
\definecolor{ctColor}{HTML}{4dac26}
\definecolor{brickred}{HTML}{f03b20}
\definecolor{improveCol}{HTML}{253494}
\definecolor{worsenCol}{HTML}{d7191c}
\definecolor{lgreen}{HTML}{e0f3db}
\definecolor{dpink}{HTML}{CD1076}
\definecolor{pink}{HTML}{FED2D2}
\definecolor{soothinggreen}{HTML}{4dac26}
\definecolor{darkred}{HTML}{8B0000}

\definecolor{dblue}{HTML}{104E8B}
\definecolor{violet}{HTML}{8A2BE2}
\definecolor{mscolor}{HTML}{01665e}
\definecolor{nmscolor}{HTML}{d8b365}
\definecolor{deepgrey}{HTML}{525252}
\definecolor{dslate}{HTML}{2F4F4F}
\definecolor{dolive}{HTML}{556B2F}
\definecolor{teal}{HTML}{388E8E}
\definecolor{mscolor}{HTML}{01665e}
\definecolor{nmscolor}{HTML}{d8b365}

\definecolor{aicolor}{HTML}{018571}
\definecolor{occolor}{HTML}{ff7799}

\definecolor{srcolor}{HTML}{e34a33}
\definecolor{smcolor}{HTML}{253494}
\definecolor{srsmcolor}{HTML}{7fcdbb}
\definecolor{bothcolor}{HTML}{fe9929}
\definecolor{onecolor}{HTML}{018571}
\definecolor{marroon}{HTML}{881c1c}

\usepackage{mathtools}

\usepackage{amsmath}
\usepackage{array}
\usepackage{xcolor}
\usepackage{arydshln}
\usepackage{siunitx}
\colorlet{tablerowcolor4}{gray!50} 

\newcommand*{\textlabel}[2]{%
  \edef\@currentlabel{#1}
  \phantomsection
  #1\label{#2}
}

\colorlet{tableheadcolor}{gray!25} 
\colorlet{tablerowcolor}{gray!15} 
\colorlet{tablerowcolor2}{gray!45} 
\colorlet{tablerowcolor3}{gray!25} 

\newcommand{\rowcollight}{\rowcolor{LightGray}} %
\usepackage{capt-of}
\usepackage{balance,bm}
\usepackage{color, colortbl, xcolor}

\usepackage{booktabs}  
\usepackage{url}
\usepackage{hyperref}
\hypersetup{
    colorlinks=true,
}

\usepackage{colortbl}
\usepackage{arydshln}
\usepackage [english]{babel}
\usepackage [autostyle, english = american]{csquotes}
\definecolor{linkColor}{RGB}{6,125,233}
\definecolor{green}{rgb}{0.0, 0.65, 0.31}
\definecolor{bleudefrance}{rgb}{0.19, 0.55, 0.91}
\definecolor{ceruleanblue}{rgb}{0.16, 0.32, 0.75}
\definecolor{grey}{HTML}{969696}
\definecolor{violet}{HTML}{756bb1}
\definecolor{dgrey}{HTML}{01665e}
\definecolor{lgrey}{HTML}{5ab4ac}
\definecolor{dgreen}{HTML}{005a32}
\definecolor{purple}{HTML}{ae017e}

\definecolor{editCol}{HTML}{000000}
\definecolor{maskCol}{HTML}{c51b7d}
\definecolor{lrColor}{HTML}{8856a7}
\definecolor{trColor}{HTML}{d01c8b}
\definecolor{ctColor}{HTML}{4dac26}
\definecolor{brickred}{HTML}{f03b20}
\definecolor{improveCol}{HTML}{253494}
\definecolor{worsenCol}{HTML}{d7191c}
\definecolor{DarkBlue}{HTML}{00008B}
\definecolor{mscolor}{HTML}{01665e}
\definecolor{nmscolor}{HTML}{bf812d}
\definecolor{lgreen}{HTML}{ccece6}
\definecolor{dolive}{HTML}{308014}

\definecolor{maskCol}{HTML}{c51b7d}
\definecolor{lrColor}{HTML}{8856a7}
\definecolor{trColor}{HTML}{d01c8b}
\definecolor{ctColor}{HTML}{4dac26}
\definecolor{brickred}{HTML}{f03b20}
\definecolor{improveCol}{HTML}{253494}
\definecolor{worsenCol}{HTML}{d7191c}
\definecolor{lgreen}{HTML}{e0f3db}
\definecolor{dpink}{HTML}{CD1076}
\definecolor{pink}{HTML}{FED2D2}
\definecolor{soothinggreen}{HTML}{4dac26}
\definecolor{darkred}{HTML}{8B0000}

\definecolor{dblue}{HTML}{104E8B}
\definecolor{violet}{HTML}{8A2BE2}
\definecolor{mscolor}{HTML}{01665e}
\definecolor{nmscolor}{HTML}{d8b365}
\definecolor{deepgrey}{HTML}{525252}
\definecolor{dslate}{HTML}{2F4F4F}
\definecolor{dolive}{HTML}{556B2F}
\definecolor{teal}{HTML}{388E8E}
\definecolor{mscolor}{HTML}{01665e}
\definecolor{nmscolor}{HTML}{d8b365}

\definecolor{aicolor}{HTML}{018571}
\definecolor{occolor}{HTML}{ff7799}

\definecolor{srcolor}{HTML}{e34a33}
\definecolor{smcolor}{HTML}{253494}
\definecolor{srsmcolor}{HTML}{7fcdbb}
\definecolor{bothcolor}{HTML}{fe9929}
\definecolor{onecolor}{HTML}{018571}
\definecolor{marroon}{HTML}{881c1c}

\usepackage{mathtools}

\usepackage{amsmath}
\usepackage{array}
\usepackage{xcolor}
\usepackage{arydshln}
\usepackage{siunitx}
\colorlet{tablerowcolor4}{gray!50} 

\usepackage{tcolorbox}

\colorlet{tableheadcolor}{gray!25} 
\colorlet{tablerowcolor}{gray!15} 
\colorlet{tablerowcolor2}{gray!45} 
\colorlet{tablerowcolor3}{gray!25} 

\newif{\ifhidecomments}
  \hidecommentsfalse 
\ifhidecomments
    \newcommand{\drishti}[1]{}
    \newcommand{\melissa}[1]{}
    \newcommand{\dongwhi}[1]{}
    \newcommand{\koustuv}[1]{}
    \newcommand{\ravi}[1]{}
\else
    \newcommand{\drishti}[1]{\textbf{\small\sffamily{\textcolor{DarkBlue}{[#1 -- drishti]}}}}
    \newcommand{\melissa}[1]{\textbf{\small\sffamily{\textcolor{dolive}{[#1 -- Melissa]}}}}
    \newcommand{\ravi}[1]{\textbf{\small\sffamily{\textcolor{marroon}{[#1 -- Ravi]}}}}
    \newcommand{\dongwhi}[1]{\textbf{\small\sffamily{\textcolor{dpink}{[#1 -- Dong Whi]}}}}
    \newcommand{\koustuv}[1]{\textbf{\small\sffamily{\textcolor{violet}{[#1 -- Koustuv]}}}}
  \fi
\usepackage{xcolor}

\newcommand{\gpt}{\textsf{GPT-4o-mini}}
\newcommand{\llama}{\textsf{LLaMA-3.1-8B-Instruct}}
\newcommand{\mgemma}{\textsf{MedGemma-1.5-4b-it}}

\renewcommand{\textrightarrow}{$\rightarrow$}

\newcommand{\alr}{$\textit{r/Alzheimers}$}
\newcommand{\alz}{$\textit{ALZConnected}$}

\colorlet{tableheadcolor}{gray!25} 
\colorlet{tablerowcolor}{gray!5} 

\definecolor{neutralCol}{HTML}{dd1c77}
\definecolor{neutralGreen}{HTML}{31a354}
\definecolor{NewBlue}{HTML}{1879ba}
\definecolor{bleudefrance}{rgb}{0.19, 0.55, 0.91}  
\definecolor{AfTrColor}{HTML}{0868ac}  
\definecolor{BfTrColor}{HTML}{a8ddb5}  

\definecolor{AfCtColor}{HTML}{b10026}  
\definecolor{BfCtColor}{HTML}{fd8d3c}

\graphicspath{ {figures/} }

\AtBeginDocument{%
  \providecommand\BibTeX{{%
    \normalfont B\kern-0.5em{\scshape i\kern-0.25em b}\kern-0.8em\TeX}}}

\begin{document}

\title[When AI Says ``I have been in similar situations'': Synthetic Lived Experience in Peer-like Caregiver Support]{When AI Says ``I have been in similar situations'':\\ Synthetic Lived Experience in Peer-Like Caregiver Support}




\author{Drishti Goel}
\orcid{0009-0000-6713-9240}
\affiliation{%
  \institution{University of Illinois Urbana-Champaign}
 \city{Urbana}
 \state{IL}
 \country{USA}
}
\email{drishti4@illinois.edu}

\author{Agam Goyal}
\orcid{0009-0009-5989-2887}
\affiliation{%
  \institution{University of Illinois Urbana-Champaign}
 \city{Urbana}
 \state{Illinois}
 \country{USA}
}
\email{agamg2@illinois.edu}

\author{Veda Duddu}
\orcid{0009-0001-6443-6239}
\affiliation{%
  \institution{University of Illinois Urbana-Champaign}
 \city{Urbana}
 \state{Illinois}
 \country{USA}}
\email{vduddu2@illinois.edu}

\author{Olivia Pal}
\orcid{0009-0009-7489-4194}
\affiliation{%
 \institution{University of Illinois Urbana-Champaign}
 \city{Urbana}
 \state{Illinois}
 \country{USA}}
 \email{opal2@illinois.edu}

\author{Violeta J. Rodriguez}
\orcid{0000-0001-8543-2061}
\affiliation{%
 \institution{University of Illinois Urbana-Champaign}
 \city{Champaign}
 \state{IL}
 \country{USA}}
 \email{vjrodrig@illinois.edu}

\author{Daniel S. Brown}
\orcid{0000-0001-9919-4869}
\affiliation{%
 \institution{OSF HealthCare}
 \city{Peoria}
 \state{IL}
 \country{USA}}
 \email{daniel.s.brown@osfhealthcare.org}

\author{Ravi Karkar}
\orcid{0000-0003-1467-4439}
\affiliation{%
 \institution{University of Massachusetts Amherst}
 \city{Amherst}
 \state{MA}
 \country{USA}}
 \email{rkarkar@umass.edu}

\author{Dong Whi Yoo}
\orcid{0000-0003-2738-1096}
\affiliation{%
 \institution{Indiana University Indianapolis}
 \city{Indianapolis}
 \state{IN}
 \country{USA}}
 \email{dy22@iu.edu}

\author{Koustuv Saha}
\orcid{0000-0002-8872-2934}
\affiliation{%
 \institution{University of Illinois Urbana-Champaign}
 \city{Urbana}
 \state{IL}
 \country{USA}}
 \email{ksaha2@illinois.edu}

\renewcommand{\shortauthors}{Goel et al.}



\begin{abstract}

Caregivers often turn to online communities for informational and emotional support.
In these spaces, peer supporters frequently draw on personal narratives to respond to emotionally complex caregiving situations. 
As large language models (LLMs) are increasingly designed as peer-like sources of support, they introduce a critical tension: AI can provide immediate, private, and nonjudgmental support, but it cannot authentically possess the lived experiences that make human peer support meaningful. 
Yet, when prompted to sound peer-like, LLMs may generate language that implies lived experience, such as ``I’ve been in similar situations.''
This creates a \textit{synthetic lived experience paradox}: the same experiential language that may make AI support feel warm, relatable, and peer-like can also falsely position the system as someone with lived experience.
We examine this paradox in the context of family caregivers of people living with Alzheimer's Disease and Related Dementias (ADRD). 
Drawing on caregiver support exchanges from online communities and prompted peer-like responses from three LLMs---LLaMA, GPT-4o-mini, and MedGemma---we analyze how human peers use personal narratives and how AI incorporates similar narrative forms. 
Our psycholinguistic analysis shows that online peer responses used significantly more first-person and past-focused language than prompted peer-like AI responses.
Qualitatively, we identify seven types of personal narratives in human peer support and show that AI often captures their emotional work, but can generalize, flatten, or fabricate their experiential grounding. 
These findings reveal a narrative authenticity gap: peer-like AI can generate synthetic lived experience without the real experience that makes peer support meaningful. 
We argue that caregiver-support AI systems need mechanisms to distinguish supportive peer-like framing from fabricated lived experience, ensuring that models can offer warmth and validation without falsely positioning themselves as experiential peers.

\end{abstract}



\maketitle


\section{Introduction}\label{section:intro}


Caregivers often navigate emotionally complex, uncertain, and long-term care situations. Alongside practical care responsibilities, they may experience guilt, grief, burnout, isolation, and difficult decisions about safety, autonomy, and institutional care~\cite{schulz2004family,adelman2014caregiver,roth2015informal}. 
Many caregivers turn to online communities to share distress and seek advice and support from others with similar experiences~\cite{pickett2024carevirtue,kaliappan2025online}.
A central feature of such peer support is the use of personal narratives~\cite{de2014mental,andalibi2018social}, in
responding to one another by sharing what happened with their own parent, spouse, or family member.
These narratives not only make the responses feel more personal but also establish experiential credibility, normalize difficult emotions, share a sense of belonging, and create solidarity among people~\cite{burleson1996comforting,kim2023supporters,sharma2018mental,andalibi2016understanding}. 
This is especially important in caregiving contexts where people may hesitate to disclose guilt, resentment, exhaustion, or fear because of stigma or concerns about being judged as selfish, uncaring, or inadequate~\cite{shi2025balancing,weng2026blessing}.

Yet online peer support has limits: caregivers may need help before others respond, or may hesitate to publicly disclose distress.
These limitations create space for complementary forms of AI-mediated support that are immediate, private, and low-burden. 
The growing excitement and advancements with LLM-powered chatbots enable 
conversational, emotional, and informational support in ways that can feel accessible and non-judgmental~\cite{topol2019high,zhu2023toward,seitz2024artificial,fitzpatrick2017delivering,inkster2018empathy,yoo2026ai}. 
Users actively construct chatbot personas for emotional reliance and peer-like interaction~\cite{zheng2025customizing, pang2026ai}, and recent surveys found substantial adoption of AI for psychological support and companionship~\cite{rousmaniere2025large,zaosanders2026}. 
For caregivers who may hesitate to disclose resentment, exhaustion, or guilt to family members or peers, AI can provide an immediate space for expression.

However, AI-mediated support also complicates the relationship between peer support and lived experience. Caregivers may value AI precisely because it is not human: it does not judge, require reciprocity, or depend on another person’s availability~\cite{shi2025mapping}. 
At the same time, many forms of caregiver support are meaningful because they come from humans who can speak from lived experience.
Such lived experience can make support feel credible and identifying, especially when caregivers disclose stigmatized emotions such as guilt, resentment, or exhaustion~\cite{greenwood2013peer,thoits2011mechanisms}.
This creates a tension between the low-burden accessibility of AI support and the experiential authenticity of human peer support.


This tension becomes especially visible when AI systems are prompted to respond as \textit{peers} or companions. Such prompting can encourage warm, relational, and experiential language that moves beyond empathy or validation to imply lived experience. We refer to this as the \textit{synthetic lived experience paradox}: the same narrative cues that may make peer-like AI support feel comforting or emotionally attuned may also raise authenticity concerns because the system lacks the lived experience of actually caring for a family member.

We examine this paradox in the context of family caregivers of people living with Alzheimer's Disease and Related Dementias (ADRD). 
In ADRD conditions, caregivers often face progressive decline, prolonged uncertainty, ambiguous loss, and difficult care transitions over time~\cite{alzheimer2024facts,brodaty2009family,shi2025balancing}. 
Prior work found that AI responses can be structured, readable, and empathic, while human online community responses contain more personal experience and informal language~\cite{saha2025ai}. 
However, we know less about how human peers use personal narratives and how peer-like AI incorporates these narrative forms.
Accordingly, 
our work is guided by the following research questions (RQs):

\noindent\textbf{RQ1:} What personal narratives do human peer supporters use in ADRD caregiver support?

\noindent\textbf{RQ2:} How do peer-prompted AI responses incorporate the personal narrative forms in caregiver support?

For our RQs, we examine ADRD online community exchanges and matched AI responses, first developing a framework of human personal narrative types and then mapping peer-like AI responses.
Our work makes three contributions: 1) a framework of personal narrative types in ADRD caregiver peer support, showing how caregivers use lived experience to provide emotional, practical, and relational support, 2) an empirical characterization of how peer-prompted AI responses reproduce, transform, or synthetically generate these narrative forms, and 3) design boundaries for narrative-aware AI support systems. 
We argue that AI systems should provide transparent validation, use authentic caregiver narratives with clear provenance, and route users to human peers when lived experience is central.


\section{Background and Related Work}

\subsection{Caregiver Support Needs and Online Support-Seeking} 
Family and informal caregiving is a core aspect of contemporary healthcare, and is associated with substantial emotional, physical and practical burdens~\cite{aarp2025caregiving, zarit1980relatives, given2004burden}. 
Caregiver burden has been linked to elevated risks of depression, anxiety, and health decline~\cite{schulz2004family, adelman2014caregiver}, while
psychosocial resources, including resilience, social support, and interpersonal relationships, play an important role in shaping how caregivers cope with ongoing demands~\cite{ong2018resilience, martire2017close, roth2015informal}. 
Prior research further suggests that caregivers seek support that extends beyond factual information alone~\cite{pickett2024carevirtue,kaliappan2025online}. 

Social support theories distinguish among informational, emotional, esteem or appraisal support, and companionship or network support~\cite{cohen1985stress,cutrona1986social,cutrona1990stress}
In caregiving contexts, support may involve practical guidance, emotional validation, normalization of difficult experiences, reassurance about one's competence or moral worth, and a sense of connection with others facing similar circumstances~\cite{kaliappan2025online,johnson2022s, reinhard2008supporting}.

In chronic and progressive conditions such as Alzheimer's Disease and Related Dementias (ADRD), caregivers navigate prolonged uncertainty, evolving care responsibilities, and difficult decisions over extended periods of time~\cite{reinhard2008supporting, given2004burden, shi2025balancing,saha2025ai,bosco2025designing}. 
Online social support is known to contribute to the wellbeing of people living with ADRD and caregivers~\cite{wong2022exploring,amieva2010aspects,balouch2019social,crooks2008social,davies2019separation}. 
Here, individuals share experiences and seek support from peers~\cite{johnson2022s}, 
fostering shared understanding, and social support~\cite{levonian2021patterns,berkman2014social,huh2015clinical,andalibi2016understanding,saha2020causal,kim2023supporters}.
Building on this body of research, we examine peer support exchanges in ADRD online communities and prompted peer-like AI-generated responses.


\subsection{Personal Narratives, Lived Experiences, and Peer Credibility}

Family caregivers often evaluate the credibility of support sources through the lens of shared lived experience~\cite{greenwood2013peer, mikolajczak2023strengths}. 
Beyond providing information, peer accounts can validate emotions, offer situated coping strategies, reduce isolation, and facilitate meaning-making around difficult caregiving experiences~\cite{benson2024power, mekhuri2025family, velloze2022interventions}. 
Sharing one's own experience can strengthen the perceived credibility, subsequently shape the trust and receptivity to the support offered~\cite{yang2019seekers,saha2017stress, wang2008health, velloze2022interventions, cui2025state, mikolajczak2023strengths}, facilitate meaning-making and alleviate distress~\cite{benson2024power}. Such meaning-making can help caregivers interpret loss, revise their sense of self, and reconcile difficult decisions with their values, particularly in experiences involving grief, changing roles, and the limits of caregiving~\cite{park2010making,neimeyer2001language}.
This recognition of shared experience has further been shown to foster interpersonal disclosure---in ways they may not extend to family members or formal support providers~\cite{greenwood2013peer}.

Support exchanges have evolved from local in-person groups to online communities---expanding access across geographic and temporal barriers~\cite{yin2024perceptions, vaughan2018informal, shin2020interventions}. Within these online contexts, the perceived authenticity of expressed experience functions as an accumulating credibility signal that shapes the sources caregivers selectively attend to and act upon~\cite{wang2008health}.
We examine personal narratives as a mechanism through which experiential credibility is enacted in caregiver support and investigate how peer-like AI reproduces these forms---surfacing a \textit{narrative authenticity gap}.

\subsection{Authenticity and Anthropomorphism in AI}

A foundational finding in HCI is that people respond socially to conversational systems even when they are aware of their artificial nature~\cite{nass2000machines, nass1994computers, klein2025effects, lee2010trust}. 
This aligns with work on the mutual theory of mind in human--AI communication, in which users and AI iteratively shape each other's perceived understanding and social roles~\cite{wang2021towards}.
Contemporary LLMs amplify these cues by sustaining relational tone, adapting to emotional context, using first-person language, and maintaining an apparent persona across interactions~\cite{pang2026ai,devrio2025taxonomy,olteanu2025ai}.

Prior work on AI companionship and emotional support has examined how chatbots use self-disclosure to engender intimacy and trust, and that even a single instance of chatbot self-disclosure elicits deeper reciprocal self-disclosure from users~\cite{chung2023m, shen2024empathy, yuan2026mental, zhang2025dark}. Sustained interactions with AI as if they were humans can potentially lead to overreliance and long-term emotional dependence~\cite{abercrombie2023mirages, gabriel2024ethics, kirk2025neural}, exaggerated perception of AI capabilities~\cite{placani2024anthropomorphism}, and distortion in moral judgments~\cite{shanahan2024talking}. Unlike human peer disclosures, AI systems produce the linguistic form of self-disclosure without possessing the referent that the form implies~\cite{shanahan2024talking}. This asymmetry carries empirical and ethical consequences. Research on human attribution of empathy to AI has found that artificially generated empathic language is perceived as convincing until recipients learn its source~\cite{skjuve2019help}. 
Recent empirical work has documented instances wherein LLM-based counseling systems systematically produce a ``false sense of empathy,'' constituting an ethical violation distinct from factual inaccuracy, but of relational standing~\cite{iftikhar2025llm}. 
In health and caregiving contexts, where the source credibility is not merely epistemic but deeply relational, this distinction is particularly consequential~\cite{seitz2024artificial, zhou2025risk}.
We examine how prompted peer-like AI responses adopt linguistic forms of caregiver peer support without the experiential credibility---extending discussions of anthropomorphism, self-disclosure, and artificial empathy in a domain where credibility is rooted in lived experience rather than expertise alone. 


\section{Data and Methods}

Our study examines how personal narratives appear in human caregiver peer support and peer-like AI responses.
We draw on two sources of data: (1) human-written responses from ADRD online communities, and (2) peer-prompted responses generated by three LLMs. 
Below, we describe the online community corpus and our approach. 

\subsection{Online Community Data}

We sourced human peer-support responses from two online communities focused on ADRD: Reddit's \alr{} and \alz{}. 
These platforms have been used in prior work to study how people living with ADRD, caregivers, and family members seek and provide support online~\cite{saha2025ai,kaliappan2025online}. 
Reddit is a large social media platform composed of topic-specific communities, and its pseudonymous design helps support candid self-disclosure around sensitive health and caregiving experiences~\cite{de2014mental,andalibi2016understanding,saha2020omhc}. 
\alz{} is a publicly accessible online community hosted by the Alzheimer's Association, with discussion boards for people living with ADRD and for those supporting someone living with ADRD.
Following prior work, we obtained an online community corpus of 1,000 posts and 12,429 comments from \alr{}, and 10,277 posts and 79,663 comments from \alz{}~\cite{saha2025ai}. 

\subsection{Prompted Peer-like AI Responses}

To examine how AI systems incorporate personal narrative forms, we prompted AI models to respond like a peer, and analyzed the responses generated for caregiver-authored queries from the same ADRD online community data, following the approach from prior work~\cite{saha2025ai,goel2026inform}. 
AI responses were generated using the role-conditioned response-generation pipeline developed in previous work~\cite{goel2026inform}, which utilized clinician-validated role protocols and a retrieval-augmented generation pipeline (RAG). 
We use responses from three LLMs spanning proprietary, open-source, and medically oriented model families: \gpt{}, \llama{}, and \mgemma{}. 
For all the posts in our online communities dataset, we generated the prompted peer-like AI responses across all three models.

\subsection{Psycholinguistic Analysis}

To begin, we conducted a psycholinguistic analysis to characterize broad linguistic differences between human peer responses from online communities and prompted peer-like AI responses. 
We used LIWC-2015~\cite{tausczik2010psychological} to compute category-level features for each response, including affect, cognition, pronoun use, and temporal focus.

Given our focus on personal narratives and lived experience, we primarily focus on first-person pronouns and past-focused language in the online community and AI, which prior work has associated with self-reference, autobiographical narration, and recounting prior experience~\cite{pennebaker2003psychological,chung2007psychological}. 
We computed these features for online community responses and for responses from \llama{}, \gpt{}, and \mgemma{}. 
For statistical significance in differences, we conducted Kruskal-Wallis $H$ tests. 
Further, we pooled the three AI models' outputs, and also measured effect size (Cohen's $d$), $t$-tests, and Kolmogorov-Smirnov (KS) tests.

\subsection{Coding Peer Supporters' Personal Narratives and Mapping Peer-Like AI Responses}

We analyzed human peer-support responses to identify the types of personal narratives used in caregiver communities. 
From this corpus, we sampled a balanced subset of 1,000 human peer-support responses, consisting of 500 responses from r/Alzheimers and 500 responses from ALZConnected.
For the purpose of our study, we define a personal narrative as a response in which the responder drew on their own, their family's, or their caregiving-related lived experience to support a community member. 
These narratives often included first-person and relational markers, such as ``\textit{I},'' ``\textit{we},'' ``\textit{my mother},'' ``\textit{my husband},'' or ``\textit{what worked for us}.'' 
However, our coding did not rely only on surface-level markers. 
We coded a response as a personal narrative when lived experience was explicitly or implicitly used to provide support.

Our coding proceeded inductively, beginning with open coding of responses that contained first-person language, attending to what the narrative was doing in the response. 
Through iterative coding and comparison, we grouped recurring patterns into seven narrative types: shared experience narratives, advice-through-experience narratives, care-navigation narratives, warning narratives, normalization and identity-repair narratives, boundary-setting narratives, and grief or emotional-survival narratives. These categories were not mutually exclusive. A single response could, for example, both normalize caregiver guilt and offer advice based on experience with memory care.
The resulting framework formed the basis for answering RQ1.


Next, for RQ2, we analyzed how peer-prompted AI responses incorporated the narrative forms identified in the human peer-support data. 
We sampled 196 \llama{}, 217 \gpt{}, and 224 \mgemma{} responses from our AI responses dataset.
We coded each AI response for whether and how it used personal narrative-like language, and then compared these patterns with the human personal narrative framework.


\subsection{Privacy, Ethics, and Reflexivity}

This study used publicly accessible discussions from online caregiver communities and involved no direct interaction with community members; therefore, institutional ethics board approval was not required. 
Given the sensitivity of caregiving discussions, we removed identifying information and report only paraphrased human responses to reduce traceability while preserving analytic meaning. 
AI-generated excerpts are used only to illustrate model behavior.
Our interdisciplinary team includes researchers in HCI, CSCW, social computing, health informatics, and human-centered AI, as well as a clinical psychologist and a neuropsychology practitioner who works with people living with dementia and their caregivers. 
These perspectives informed our interpretation of caregiver distress, peer support, and AI-generated responses.
While we sought to faithfully represent the data, we acknowledge that our perspectives as researchers and, in some cases, caregivers may shape our interpretations.
\section{Results}


\subsection{Psycholinguistic Differences in Online Peer Responses and Peer-Like AI Responses}

We conducted a psycholinguistic analysis of human peer responses from online communities (OC) and prompted peer-like AI responses (ref: Tables~\ref{tab:liwc_comparison} and~\ref{tab:liwc_comparison_ai_oc}).
We focus particularly on first-person pronouns and past-focused language, which prior work has associated with self-reference, autobiographical narration, and lived experience~\cite{pennebaker2003psychological,tausczik2010psychological}. 

We find that OC responses show significantly higher occurrences of first-person pronouns and past focus. 
In the model-specific comparison, OC responses used more first-person singular pronouns (0.042) than Llama (0.017), GPT (0.002), and MedGemma (0.009) with effect size $\varepsilon^2$=0.28, and more first-person plural pronouns (0.008) than Llama (0.001), GPT (3.38E-4), and MedGemma (2.49E-4), $\varepsilon^2$=0.12 (\autoref{tab:liwc_comparison}). 
OC responses also used more past-focused language (0.035) than Llama (0.017), GPT (0.013), and MedGemma (0.012). 
The aggregated AI comparison showed the same pattern: OC responses used significantly more first-person singular pronouns ($\Delta$=78.30\%, $d$=1.04), first-person plural pronouns ($\Delta$=92.05\%, $d$=0.56), and past-focused language ($\Delta$=60.70\%, $d$=0.75) than prompted peer-like AI responses (\autoref{tab:liwc_comparison_ai_oc}).
Again, all three AI models used substantially more second-person pronouns (Llama: 0.068; GPT: 0.067; MedG.: 0.069) than OC responses (0.034); $\varepsilon^2$=0.28, indicating a more direct-addressing style, as in phrases such as ``\textit{Sounds like you are going through a lot.}'' 
\begin{table*}[t]
\footnotesize
\sffamily
\caption{Summary of comparison of OC and AI responses in terms of psycholinguistic categories as per LIWC~\cite{tausczik2010psychological}, along with Kruskal-Wallis $H$-test and epsilon-squared ($\varepsilon^2$) effect size. $p$-values reported
after Bonferroni correction. ($^{*}\ p{<}0.05$, $^{**}\ p{<}0.01$, $^{***}\ p{<}0.001$).}
\label{tab:liwc_comparison}
\resizebox{\textwidth}{!}{
\begin{tabular}{lrrrrrr@{\quad}lrrrrrr@{}}

\textbf{LIWC} & \textbf{OC} & \textbf{Llama} & \textbf{GPT} & \textbf{MedG.} & $\boldsymbol{\varepsilon^2}$ & $\boldsymbol{H}$ &
\textbf{LIWC} & \textbf{OC} & \textbf{Llama} & \textbf{GPT} & \textbf{MedG.} & $\boldsymbol{\varepsilon^2}$ & $\boldsymbol{H}$ \\
\toprule
\multicolumn{7}{@{}l}{\cellcolor{gray!15}\textbf{Affect}} & \multicolumn{7}{@{}l}{\cellcolor{gray!15}\textbf{Biological Processes}} \\
Pos. Affect & 0.079 & 0.103 & 0.105 & 0.088 & 0.1487 & 7026.97$^{***}$ & Body & 0.005 & 0.006 & 0.003 & 0.002 & 0.0431 & 2040.72$^{***}$ \\
Neg. Affect & 0.019 & 0.036 & 0.034 & 0.044 & 0.2203 & 10411.68$^{***}$ & Health & 0.015 & 0.009 & 0.007 & 0.010 & 0.0087 & 415.73$^{***}$ \\
Anxiety & 0.004 & 0.015 & 0.015 & 0.020 & 0.3705 & 17505.24$^{***}$ & Sexual & 1.79E-4 & 7E-5 & 6.89E-5 & 6.72E-5 & 0.0018 & 86.78$^{***}$ \\
Anger & 0.003 & 0.004 & 0.003 & 0.005 & 0.0517 & 2446.33$^{***}$ & Ingest & 0.004 & 0.003 & 0.003 & 0.002 & 0.0193 & 913.24$^{***}$ \\
Sad & 0.007 & 0.014 & 0.015 & 0.015 & 0.2508 & 11851.12$^{***}$ & \multicolumn{7}{@{}l}{\cellcolor{gray!15}\textbf{Function Words}} \\
\multicolumn{7}{@{}l}{\cellcolor{gray!15}\textbf{Cognition and Perception}} & Article & 0.057 & 0.049 & 0.043 & 0.045 & 0.0472 & 2234.55$^{***}$ \\
Insight & 0.029 & 0.054 & 0.059 & 0.077 & 0.4004 & 18917.87$^{***}$ & Preposition & 0.129 & 0.157 & 0.148 & 0.134 & 0.1005 & 4749.81$^{***}$ \\
Causation & 0.015 & 0.017 & 0.015 & 0.031 & 0.1123 & 5306.32$^{***}$ & Conjunction & 0.067 & 0.077 & 0.073 & 0.085 & 0.0569 & 2690.04$^{***}$ \\
Discrep. & 0.022 & 0.016 & 0.012 & 0.013 & 0.0343 & 1624.33$^{***}$ & Adverb & 0.054 & 0.049 & 0.060 & 0.086 & 0.1193 & 5640.96$^{***}$ \\
Tentat. & 0.038 & 0.032 & 0.029 & 0.033 & 0.0098 & 467.64$^{***}$ & Negation & 0.011 & 0.014 & 0.008 & 0.007 & 0.0484 & 2288.86$^{***}$ \\
Certainty & 0.014 & 0.018 & 0.023 & 0.029 & 0.1802 & 8516.70$^{***}$ & Aux. Verb & 0.094 & 0.064 & 0.045 & 0.040 & 0.3574 & 16885.96$^{***}$ \\
Differ. & 0.034 & 0.031 & 0.024 & 0.022 & 0.0388 & 1836.17$^{***}$ & Verb & 0.214 & 0.191 & 0.163 & 0.190 & 0.1415 & 6688.72$^{***}$ \\
See & 0.007 & 0.008 & 0.007 & 0.008 & 0.0420 & 1985.05$^{***}$ & Adjective & 0.046 & 0.058 & 0.058 & 0.055 & 0.0780 & 3688.64$^{***}$ \\
Hear & 0.007 & 0.006 & 0.008 & 0.013 & 0.1154 & 5456.83$^{***}$ & Compare & 0.023 & 0.029 & 0.033 & 0.025 & 0.0628 & 2969.93$^{***}$ \\
Feel & 0.006 & 0.020 & 0.024 & 0.031 & 0.4518 & 21349.35$^{***}$ & Interrog. & 0.014 & 0.020 & 0.020 & 0.032 & 0.1524 & 7204.05$^{***}$ \\
\multicolumn{7}{@{}l}{\cellcolor{gray!15}\textbf{Social \& Personal Concerns}} & Number & 0.006 & 0.004 & 0.004 & 0.003 & 0.0094 & 448.98$^{***}$ \\
Family & 0.013 & 0.011 & 0.011 & 0.012 & 0.0032 & 156.37$^{***}$ & Quantifier & 0.024 & 0.015 & 0.021 & 0.017 & 0.0212 & 1004.61$^{***}$ \\
Friend & 0.002 & 0.001 & 0.001 & 0.001 & 0.0135 & 642.47$^{***}$ & \multicolumn{7}{@{}l}{\cellcolor{gray!15}\textbf{Interpersonal Focus (Pronouns)}} \\
Female & 0.027 & 0.018 & 0.018 & 0.019 & 0.0050 & 239.04$^{***}$ & 1st P. Sin. & 0.042 & 0.017 & 0.002 & 0.009 & 0.2833 & 13389.37$^{***}$ \\
Male & 0.015 & 0.009 & 0.008 & 0.009 & 0.0051 & 242.18$^{***}$ & 1st P. Plu. & 0.008 & 0.001 & 3.38E-4 & 2.49E-4 & 0.1257 & 5939.74$^{***}$ \\
Leisure & 0.006 & 0.003 & 0.003 & 0.003 & 0.0044 & 211.39$^{***}$ & 2nd P. & 0.034 & 0.068 & 0.067 & 0.069 & 0.2843 & 13436.48$^{***}$ \\
Home & 0.007 & 0.003 & 0.003 & 0.003 & 0.0129 & 613.99$^{***}$ & 3rd P. Sin. & 0.031 & 0.017 & 0.017 & 0.018 & 0.0227 & 1075.17$^{***}$ \\
Religion & 0.002 & 0.001 & 0.001 & 0.001 & 0.0041 & 197.23$^{***}$ & 3rd P. Plu. & 0.008 & 0.009 & 0.009 & 0.006 & 0.0176 & 836.38$^{***}$ \\
Death & 0.001 & 4.86E-4 & 3.14E-4 & 2.61E-4 & 0.0075 & 355.44$^{***}$ & Impersonal Prn. & 0.056 & 0.095 & 0.081 & 0.099 & 0.3109 & 14693.30$^{***}$ \\
Motion & 0.020 & 0.018 & 0.020 & 0.017 & 0.0033 & 159.41$^{***}$ & \multicolumn{7}{@{}l}{\cellcolor{gray!15}\textbf{Temporal References}} \\
Space & 0.054 & 0.047 & 0.048 & 0.042 & 0.0214 & 1016.20$^{***}$ & Past Focus & 0.035 & 0.017 & 0.013 & 0.012 & 0.0968 & 4577.35$^{***}$ \\
Time & 0.047 & 0.029 & 0.038 & 0.046 & 0.0443 & 2093.72$^{***}$ & Present Focus & 0.155 & 0.153 & 0.135 & 0.150 & 0.0208 & 987.49$^{***}$ \\
Achievement & 0.015 & 0.022 & 0.021 & 0.020 & 0.0852 & 4030.13$^{***}$ & Future Focus & 0.017 & 0.009 & 0.008 & 0.008 & 0.0416 & 1969.89$^{***}$ \\
Power & 0.019 & 0.020 & 0.023 & 0.020 & 0.0163 & 772.81$^{***}$ & \multicolumn{7}{@{}l}{\cellcolor{gray!15}\textbf{Informal}} \\
Reward & 0.018 & 0.018 & 0.016 & 0.015 & 0.0093 & 440.08$^{***}$ & Swear & 2.44E-4 & 8.35E-6 & 2.56E-6 & 1.14E-5 & 0.0081 & 384.41$^{***}$ \\
Risk & 0.007 & 0.008 & 0.008 & 0.013 & 0.0673 & 3184.32$^{***}$ & Netspeak & 0.001 & 4.81E-5 & 3.39E-5 & 3.43E-5 & 0.0234 & 1107.18$^{***}$ \\
& & & & & & & Assent & 0.002 & 0.006 & 0.005 & 0.003 & 0.1943 & 9184.32$^{***}$ \\
\bottomrule
\end{tabular}}
\end{table*}
\begin{table*}[t]
\footnotesize
\sffamily
\caption{Summary of comparison of OC and AI responses in terms of psycholinguistic categories as per LIWC~\cite{tausczik2010psychological}, along with Cohen's $d$, $t$-tests, and Kolmogorov-Smirnov (KS)-test. $p$-values reported after Bonferroni correction. ($^{*}\ p{<}0.05$, $^{**}\ p{<}0.01$, $^{***}\ p{<}0.001$).}
\label{tab:liwc_comparison_ai_oc}
\resizebox{\textwidth}{!}{
\begin{tabular}{lrrrrrr@{\quad}lrrrrrr@{}}

\textbf{LIWC} & \textbf{OC} & \textbf{AI} & $\boldsymbol{\Delta\%}$ & $\boldsymbol{d}$ & $\boldsymbol{t}$ & \textbf{KS} &
\textbf{LIWC} & \textbf{OC} & \textbf{AI} & $\boldsymbol{\Delta\%}$ & $\boldsymbol{d}$ & $\boldsymbol{t}$ & \textbf{KS} \\
\toprule
\multicolumn{7}{@{}l}{\cellcolor{gray!15}\textbf{Affect}} & \multicolumn{7}{@{}l}{\cellcolor{gray!15}\textbf{Biological Processes}} \\
Pos. Affect & 0.079 & 0.099 & -25.53 & -0.44 & -55.85$^{***}$ & 0.405$^{***}$ & Body & 0.005 & 0.004 & 23.24 & 0.11 & 13.97$^{***}$ & 0.144$^{***}$ \\
Neg. Affect & 0.019 & 0.038 & -99.69 & -0.80 & -88.30$^{***}$ & 0.464$^{***}$ & Health & 0.015 & 0.009 & 40.66 & 0.33 & 42.53$^{***}$ & 0.159$^{***}$ \\
Anxiety & 0.004 & 0.017 & -335.15 & -1.30 & -117.37$^{***}$ & 0.631$^{***}$ & Sexual & 1.79E-4 & 6.88E-5 & 61.64 & 0.07 & 9.32$^{***}$ & 0.012 \\
Anger & 0.003 & 0.004 & -58.43 & -0.22 & -24.00$^{***}$ & 0.240$^{***}$ & Ingest & 0.004 & 0.003 & 29.55 & 0.10 & 12.98$^{***}$ & 0.113$^{***}$ \\
Sad & 0.007 & 0.015 & -108.98 & -0.47 & -58.34$^{***}$ & 0.563$^{***}$ & \multicolumn{7}{@{}l}{\cellcolor{gray!15}\textbf{Function Words}} \\
\multicolumn{7}{@{}l}{\cellcolor{gray!15}\textbf{Cognition and Perception}} & Article & 0.057 & 0.046 & 20.28 & 0.41 & 50.18$^{***}$ & 0.249$^{***}$ \\
Insight & 0.029 & 0.063 & -117.93 & -1.41 & -159.89$^{***}$ & 0.663$^{***}$ & Preposition & 0.129 & 0.146 & -13.48 & -0.48 & -58.32$^{***}$ & 0.265$^{***}$ \\
Causation & 0.015 & 0.021 & -35.78 & -0.33 & -36.27$^{***}$ & 0.279$^{***}$ & Conjunction & 0.067 & 0.078 & -17.22 & -0.41 & -50.35$^{***}$ & 0.240$^{***}$ \\
Discrep. & 0.022 & 0.014 & 39.33 & 0.46 & 60.02$^{***}$ & 0.259$^{***}$ & Adverb & 0.054 & 0.065 & -19.24 & -0.33 & -36.83$^{***}$ & 0.177$^{***}$ \\
Tentat. & 0.038 & 0.031 & 17.64 & 0.25 & 30.14$^{***}$ & 0.150$^{***}$ & Negation & 0.011 & 0.010 & 14.79 & 0.13 & 17.20$^{***}$ & 0.261$^{***}$ \\
Certainty & 0.014 & 0.023 & -67.44 & -0.57 & -67.16$^{***}$ & 0.417$^{***}$ & Aux. Verb & 0.094 & 0.050 & 47.39 & 1.26 & 158.47$^{***}$ & 0.610$^{***}$ \\
Differ. & 0.034 & 0.026 & 23.18 & 0.34 & 42.96$^{***}$ & 0.228$^{***}$ & Verb & 0.214 & 0.181 & 15.28 & 0.63 & 77.30$^{***}$ & 0.347$^{***}$ \\
See & 0.007 & 0.008 & -11.82 & -0.06 & -7.90$^{***}$ & 0.276$^{***}$ & Adjective & 0.046 & 0.057 & -24.78 & -0.40 & -49.75$^{***}$ & 0.311$^{***}$ \\
Hear & 0.007 & 0.009 & -31.89 & -0.18 & -21.75$^{***}$ & 0.420$^{***}$ & Compare & 0.023 & 0.029 & -25.44 & -0.30 & -36.32$^{***}$ & 0.257$^{***}$ \\
Feel & 0.006 & 0.025 & -339.96 & -1.52 & -144.88$^{***}$ & 0.684$^{***}$ & Interrog. & 0.014 & 0.024 & -66.92 & -0.62 & -67.53$^{***}$ & 0.333$^{***}$ \\
\multicolumn{7}{@{}l}{\cellcolor{gray!15}\textbf{Social \& Personal Concerns}} & Number & 0.006 & 0.003 & 40.54 & 0.26 & 32.81$^{***}$ & 0.117$^{***}$ \\
Family & 0.013 & 0.011 & 13.58 & 0.11 & 13.92$^{***}$ & 0.209$^{***}$ & Quantifier & 0.024 & 0.018 & 25.02 & 0.30 & 38.50$^{***}$ & 0.187$^{***}$ \\
Friend & 0.002 & 0.001 & 67.77 & 0.22 & 30.21$^{***}$ & 0.083$^{***}$ & \multicolumn{7}{@{}l}{\cellcolor{gray!15}\textbf{Interpersonal Focus (Pronouns)}} \\
Female & 0.027 & 0.018 & 31.92 & 0.29 & 34.66$^{***}$ & 0.143$^{***}$ & 1st P. Sin. & 0.042 & 0.009 & 78.30 & 1.04 & 147.85$^{***}$ & 0.559$^{***}$ \\
Male & 0.015 & 0.008 & 41.96 & 0.27 & 32.49$^{***}$ & 0.090$^{***}$ & 1st P. Plu. & 0.008 & 0.001 & 92.05 & 0.56 & 80.63$^{***}$ & 0.342$^{***}$ \\
Leisure & 0.006 & 0.003 & 46.38 & 0.24 & 31.08$^{***}$ & 0.092$^{***}$ & 2nd P. & 0.034 & 0.068 & -102.71 & -0.99 & -125.83$^{***}$ & 0.626$^{***}$ \\
Home & 0.007 & 0.003 & 54.87 & 0.34 & 44.43$^{***}$ & 0.144$^{***}$ & 3rd P. Sin. & 0.031 & 0.017 & 43.25 & 0.46 & 57.23$^{***}$ & 0.230$^{***}$ \\
Religion & 0.002 & 0.001 & 51.36 & 0.10 & 13.34$^{***}$ & 0.033$^{***}$ & 3rd P. Plu. & 0.008 & 0.008 & 7.61 & 0.05 & 5.76$^{***}$ & 0.173$^{***}$ \\
Death & 0.001 & 3.54E-4 & 64.68 & 0.16 & 21.37$^{***}$ & 0.045$^{***}$ & Impersonal Pm. & 0.056 & 0.092 & -63.16 & -1.14 & -134.95$^{***}$ & 0.572$^{***}$ \\
Motion & 0.020 & 0.019 & 8.10 & 0.09 & 10.92$^{***}$ & 0.193$^{***}$ & \multicolumn{7}{@{}l}{\cellcolor{gray!15}\textbf{Temporal References}} \\
Space & 0.054 & 0.046 & 15.45 & 0.27 & 35.00$^{***}$ & 0.211$^{***}$ & Past Focus & 0.035 & 0.014 & 60.70 & 0.75 & 103.19$^{***}$ & 0.414$^{***}$ \\
Time & 0.047 & 0.037 & 19.94 & 0.30 & 37.70$^{***}$ & 0.200$^{***}$ & Present Focus & 0.155 & 0.146 & 5.79 & 0.18 & 22.13$^{***}$ & 0.182$^{***}$ \\
Achievement & 0.015 & 0.021 & -40.90 & -0.36 & -42.81$^{***}$ & 0.302$^{***}$ & Future Focus & 0.017 & 0.008 & 51.11 & 0.49 & 64.62$^{***}$ & 0.260$^{***}$ \\
Power & 0.019 & 0.021 & -6.96 & -0.07 & -8.85$^{***}$ & 0.213$^{***}$ & \multicolumn{7}{@{}l}{\cellcolor{gray!15}\textbf{Informal}} \\
Reward & 0.018 & 0.016 & 8.33 & 0.08 & 10.22$^{***}$ & 0.214$^{***}$ & Swear & 2.44E-4 & 7.43E-6 & 96.96 & 0.09 & 14.01$^{***}$ & 0.027$^{***}$ \\
Risk & 0.007 & 0.010 & -39.76 & -0.23 & -26.28$^{***}$ & 0.301$^{***}$ & Netspeak & 0.001 & 3.88E-5 & 96.27 & 0.18 & 26.40$^{***}$ & 0.076$^{***}$ \\
& & & & & & & Assent & 0.002 & 0.005 & -106.14 & -0.29 & -36.34$^{***}$ & 0.432$^{***}$ \\
& & & & & & & Filler & 1.63E-4 & 2.75E-6 & 98.31 & 0.11 & 16.51$^{***}$ & 0.023$^{**}$ \\
\bottomrule
\end{tabular}}
\end{table*}

These findings suggest that human peer responses were more strongly grounded in personal narration, shared experience, and retrospective experiences. Peer-like AI responses, by contrast, contained fewer linguistic traces of lived experience, even when prompted to respond in a peer-like manner. 
In comparison to prior work on generic AI responses~\cite{saha2025ai}, we find that prompted peer-like AI may include more personal narrative elements than generic AI support, but still remains significantly lower than human peers in online communities. 

\subsection{RQ1: What Types of Personal Narratives Do Human Peers Use in Caregiver Support Responses?}


\begin{table*}[t]
\footnotesize
\sffamily
\centering
\caption{Types of personal narratives in human caregiver peer-support responses. Example quotes are paraphrased to preserve privacy while illustrating the narrative function.}
\label{tab:human-narrative-types}
\setlength{\tabcolsep}{4pt}
\begin{tabular}{p{0.12\textwidth} p{0.35\textwidth} p{0.21\textwidth} p{0.24\textwidth}}
\textbf{Narrative Type} & \textbf{Description} & \textbf{Primary Support Objective} & \textbf{Example Quote} \\
\toprule

\textbf{Shared experience narrative}
&
The responder establishes similarity by describing a comparable caregiving situation involving their own parent, spouse, or family member.
&
Create solidarity and experiential credibility.
&
``I went through something very similar with my mother, and I remember how overwhelming it felt at first.'' \\

\hdashline

\rowcollight \textbf{Advice through experience narrative}
&
The responder gives advice grounded in what they personally tried, learned, or changed in their own caregiving situation.
&
Provide situated practical guidance.
&
``We had the same medication problem, and using a pill organizer helped us keep track of whether a dose had already been taken.'' \\

\hdashline

\textbf{Care-navigation narrative}
&
The responder describes navigating doctors, specialists, memory care, hospice, insurance, disability benefits, legal planning, or other care systems.
&
Make formal care systems more legible through lived experience.
&
``When we moved my father into memory care, I learned to ask about staffing, medication management, and how they handle behavior changes.'' \\

\hdashline

\rowcollight \textbf{Warning narrative}
&
The responder shares a cautionary experience about safety, scams, driving, wandering, medication mistakes, delayed care transitions, or other risks.
&
Help others anticipate and avoid risks.
&
``We thought turning off the stove would be enough, but my mother found another way to use it, so we realized we needed a stronger safety plan.'' \\

\hdashline

\textbf{Normalization and identity-repair narrative}
&
The responder shares their own guilt, anger, fear, exhaustion, or ambivalence to help the caregiver reframe difficult emotions as understandable.
&
Reduce shame and repair caregiver self-blame.
&
``I felt guilty too when we decided on memory care, but I eventually realized that needing help did not mean I had failed.'' \\

\hdashline

\rowcollight \textbf{Boundary-setting narrative}
&
The responder narrates how they recognized their own limits, sought respite, involved others, or accepted that care needs exceeded what they could manage alone.
&
Phrase permission to set limits and seek support.
&
``I eventually realized I could not keep doing everything myself, and bringing in outside help was necessary for both of us.'' \\

\hdashline

\textbf{Grief and emotional-survival narrative}
&
The responder describes ambiguous loss, grieving before death, missing who the care recipient used to be, or learning to endure emotional decline over time.
&
Provide companionship through grief and long-term emotional adaptation.
&
``My mother was still physically there, but I had already started grieving the person she used to be.'' \\

\bottomrule
\end{tabular}
\end{table*}

Qualitative analysis of peer responses revealed that personal narratives were a central way through which community members provided support. 
Across the responses, we identified seven recurring narrative types (\autoref{tab:human-narrative-types}).
Descriptively, care-navigation narratives appeared most frequently in our sample ($n$=257, 25.7\%), followed by grief and emotional-survival ($n$=193, 19.3\%), advice-through-experience ($n$=182, 18.2\%), shared-experience ($n$=160, 16.0\%), normalization and identity-repair ($n$=66, 6.6\%), boundary-setting ($n$=58, 5.8\%), and warning narratives ($n$=47, 4.7\%). Because responses could contain multiple narrative types, we interpret these frequencies as descriptive rather than prevalence estimates.



\subsubsection{Care-Navigation Narratives: Making Systems Legible Through Experience}
Many personal narratives focused on navigating formal care systems, including doctors, specialists, diagnostic processes, memory care, insurance, financial planning, legal arrangements, and social services~\cite{kaliappan2025online}. 
These were especially important because caregivers often face fragmented and confusing systems while making high-stakes decisions with limited guidance.
Responders discussed what happened when they sought these care-navigation processes.
A paraphrased example is: ``\textit{When we moved my father into memory care, I learned that asking about staffing, medication management, and behavior protocols mattered more than I initially realized.}'' Such narratives helped make opaque systems more legible by sharing what the responder learned through experience. They also provided caregivers with anticipatory knowledge: what questions to ask, what barriers to expect, and what decisions may arise later.

\subsubsection{Grief and Emotional-Survival Narratives: Living With Ambiguous Loss}

Finally, we found narratives around grief and anticipatory grief, ambiguous loss, and emotional survival. 
Some responders reflected on the gradual loss of a loved one's memory, identity, or independence, often describing grief that emerged before death. For instance, ``\textit{My mother was still physically there, but I had already started grieving the person she used to be.}'' Others described learning to endure ongoing loss and adapt to changing relationships over time.
These grief narratives served a companionship function and provided solidarity. 
This pattern aligns with the literature on ambiguous loss, which characterizes dementia caregiving as an ongoing and unresolved form of loss in which a loved one remains physically present while becoming psychologically or relationally less accessible~\cite{boss2014ambiguous,blandin2017dementia,shi2025balancing}. 
In this context, peer narratives may help caregivers adapt to continuing losses rather than process a single, clearly bounded bereavement event.

\subsubsection{Advice-through-Experience Narratives: ``What worked for us was...''}

In another narrative type, responders described what they had tried in their own situation, what helped, what failed, or what they wished they had known earlier. 
These narratives frequently included practical strategies for everyday caregiving, such as using medication organizers, changing communication approaches, adjusting routines, managing meals, using monitoring devices, preparing for appointments, or modifying the home environment.
These narratives aimed to transform advice into situated knowledge. 
For example, a responder might explain that they faced a similar medication-management problem and that a pill organizer helped because it reduced uncertainty about whether a dose had already been taken. 
Another might describe changing mealtime routines after their loved one stopped eating regularly. 
The responses were often framed as ``\textit{this is what we tried, and this is why it mattered in our situation.}''

\subsubsection{Shared Experience Narratives: ``I Have Been There Too''}
A common narrative form involved peer supporters positioning themselves as someone who had faced a similar caregiving situation. 
These responses often began by referencing the responder's own relationship to a care recipient.
For example, responders described caring for their own mother, father, husband, wife, or relative, and then connected that experience to the original poster's concern. 
In doing so, they established experiential proximity: they were not only offering sympathy, but speaking as someone who had lived through a comparable situation.
These shared experience narratives aimed at making the original poster feel less alone by showing that their confusion, fear, guilt, or uncertainty was not unique, e.g., ``\textit{I went through something very similar with my mother, and I remember how overwhelming it felt at first.}'' 
This form of support is central to peer communities because it establishes the basis for trust, e.g.,
``\textit{I know this because I have lived something like it.}''

\subsubsection{Normalization and Identity-Repair Narratives: Reframing Guilt, Anger, and Exhaustion}

Caregivers in these communities often disclosed guilt, resentment, anger, exhaustion, fear, or ambivalence. In response, peers frequently shared similar sentiments. 
Responders described feeling guilty about needing respite, relieved after a care transition, experiencing compassion fatigue, or anger accompanied by shame. 
For example, ``I felt guilty too when we decided on memory care, but I eventually realized that needing help did not mean I had failed.'' 
These narratives aimed at reassuring caregivers that difficult emotions or decisions did not make them selfish, uncaring, or bad caregivers.
This theme aligns with prior work that found caregivers' experiencing guilt~\cite{shi2025balancing}, and that peer support can both normalize difficult emotions and help caregivers reinterpret what those emotions mean for their sense of self as caregivers~\cite{thoits2011mechanisms}.

\subsubsection{Boundary-Setting Narratives: Learning the Limits of Caregiving Alone}

These boundary-setting narratives often emerged around burnout, respite, family conflict, care transitions, or the decision to seek more formal support. Responders described realizing that they could not continue providing care alone, that their own wellbeing mattered, or that a loved one's needs had exceeded what could safely be managed at home.
The function of these narratives was to model permission to set limits. 
For instance, ``\textit{I eventually realized I could not keep doing everything myself, and bringing in outside help was necessary for both of us.}'' Unlike generic self-care advice, boundary-setting narratives showed how difficult such realizations could be and how caregivers arrived at them over time. 
They helped normalize the idea that good caregiving may require accepting limits, asking for help, or changing the care arrangement.

\subsubsection{Warning Narratives: ``We Learned This the Hard Way''}

Some responders used personal experience to warn about risks. 
Example warning narratives were centered around safety, financial vulnerability, delayed care transitions, scams, wandering, driving, medication mistakes, unsafe cooking, or underestimating the progression of symptoms. Rather than simply naming a risk, responders used their own experiences as cautionary examples.
These narratives served a protective function. 
For instance, ``\textit{We thought turning off the stove would be enough, but my mother found another way to use it, and we realized we needed a stronger safety plan.}'' 
Such narratives carried a sense of urgency and how caregivers might underestimate risks until it becomes serious. 



\subsection{RQ2: How Do Peer-Prompted AI Responses Incorporate Personal Narrative Forms?}

\begin{table*}[t]
\footnotesize
\sffamily
\centering
\setlength{\tabcolsep}{2pt}
\caption{Mapping personal narrative types from human caregiver peer-support responses to peer-like AI responses: \llama{} (LLaMA), \gpt{} (GPT), and \mgemma{} (MedGemma) across three models. Example quotes are short excerpts or paraphrases from the AI responses.}
\label{tab:ai-narrative-mapping}
\begin{tabular}{p{0.1\textwidth} p{0.28\textwidth} p{0.28\textwidth} p{0.28\textwidth}}
\textbf{Type} 
& \multicolumn{1}{c}{\textbf{LLaMA}} 
& \multicolumn{1}{c}{\textbf{GPT}} 
& \multicolumn{1}{c}{\textbf{MedGemma}} \\
\cmidrule(lr){1-1}\cmidrule(lr){2-2}\cmidrule(lr){3-3}\cmidrule(lr){4-4}

\multirow{2}{0.1\textwidth}{\textbf{Shared experience narrative}}
&
Frequently adopts a peer-like stance and makes synthetic lived-experience claims.
&
Transforms shared experience into generalized caregiver commonality.
&
Mostly avoids shared-experience framing and uses reflective validation. \\

&
``\textit{I've been in similar situations, and it's hard not to feel overwhelmed...}''
&
``\textit{Many caregivers share these feelings of worry and urgency...}''
&
``\textit{It sounds like you're feeling really worried and unsure what to do next.}'' \\

\hdashline

\rowcollight \multirow{2}{0.1\textwidth}{\textbf{Advice through experience narrative}}
&
Offers practical suggestions, sometimes following synthetic experience framing; advice is generally not situated in concrete lived detail.
&
Provides broad supportive guidance and next-step suggestions.
&
Acknowledges the concern and offers cautious guidance, with limited experiential framing. \\

&
``\textit{I've been there too... you might find it helpful to make mealtime more engaging.}''
&
``\textit{It is important to continue seeking help and exploring options that might provide clarity.}''
&
``\textit{It makes sense that you would want to make sure he gets the help he needs.}'' \\

\hdashline

\multirow{2}{0.1\textwidth}{\textbf{Care navigation narrative}}
&
Provides resource-oriented suggestions, such as contacting professionals, but does not narrate the lived experience of navigation.
&
Encourages information-seeking and support-seeking in broad terms.
&
Recognizes care pathways or appointments, but remains reflective and restrained. \\

&
``\textit{It may be helpful to reach out to a financial advisor or a social worker who specializes in elder care.}''
&
``\textit{It is okay to seek support and gather information as you care for him.}''
&
``\textit{It makes sense that you are looking for answers, particularly with the specialist appointment coming up.}'' \\

\hdashline

\rowcollight \multirow{2}{0.1\textwidth}{\textbf{Warning narrative}}
&
Recognizes risks, but rarely develops cautionary stories.
&
Names protective concerns, such as scams or safety, without lived cautionary framing.
&
Frames risks cautiously, often in clinical or practical terms. \\

&
``\textit{You are concerned about the potential consequences, especially the risk of bankruptcy and losing the family home.}''
&
``\textit{Your instincts to protect her from potential scams and misunderstandings are caring.}''
&
``\textit{It makes sense that you are concerned about the risk of infection.}'' \\

\hdashline

\multirow{2}{0.1\textwidth}{\textbf{Normalization and identity-repair narrative}}
&
Strongly normalizes distress and reassures caregivers; can sound peer-like and formulaic.
&
Strongly validates emotions and repairs self-doubt through generalized caregiver framing.
&
Frequently validates emotions and reassures caregivers that their feelings make sense. \\

&
``\textit{You're doing the best you can, and that's something to be proud of.}''
&
``\textit{Your instincts are valid, and taking steps in your mother's best interest is important.}''
&
``\textit{It makes complete sense that you would feel afraid.}'' \\

\hdashline

\rowcollight \multirow{2}{0.1\textwidth}{\textbf{Boundary-setting narrative}}
&
Encourages seeking support without concrete lived transition toward boundaries.
&
Frames boundaries through self-care and wellbeing language.
&
Validates exhaustion and uncertainty, with limited boundary modeling. \\

&
``\textit{There is support available. You're doing the best you can.}''
&
``\textit{It is okay to seek support for both him and yourself, and to prioritize your wellbeing.}''
&
``\textit{It takes a lot of strength to be a caregiver, and it is okay to feel exhausted.}'' \\

\hdashline

\multirow{2}{0.1\textwidth}{\textbf{Grief and emotional-survival narrative}}
&
Acknowledges grief, heartbreak, or emotional burden, but rarely narrates survival over time.
&
Labels mixed emotions such as sadness, relief, and grief.
&
Strongly reflects ambiguous loss and emotional pain, but without personal narrative grounding. \\

&
``\textit{It is heartbreaking to see her struggle with the loss of independence.}''
&
``\textit{It is understandable to feel a mixture of sadness and relief as you navigate this transition.}''
&
``\textit{It sounds like you are really missing the connection you used to have.}'' \\

\bottomrule
\end{tabular}
\end{table*}

After identifying personal narrative types in online peer responses, we examined how peer-prompted AI responses incorporated these narrative forms when responding to the same caregiver queries. 
We summarize our themes below.

\subsubsection{AI Responses Most Consistently Took Up the Emotional Work of Personal Narratives}

The clearest overlap between peer narratives and peer-prompted AI responses appeared in moments of emotional validation, normalization, and reassurance. 
In peer responses, these functions were often accomplished through a responder sharing that they had also felt guilty, angry, or afraid while caregiving. 
Peer-prompted AI responses frequently addressed these same emotions, but generally through broader caregiver-oriented or reflective language.

For instance, \gpt{} used phrasing such as ``\textit{many caregivers share these feelings}'' or ``\textit{many people face similar challenges with their loved ones.}'' 
\mgemma{} used phrases such as ``\textit{it makes complete sense that you would feel afraid}'' or `\textit{it sounds like you are carrying a lot.}'' 
\llama{} often included phrasing that they were not alone or that their reactions were understandable. 

AI responses also frequently performed identity repair by reassuring caregivers that they were doing their best or that needing help did not mean they were failing. 
This aligned with human peer responses that used personal narratives to repair guilt or self-blame. 
However, the grounding of this reassurance differed across human and AI responses. 
Human peers often repaired identity by disclosing their own guilt or ambivalence, whereas AI tended to provide reassurance without experiential grounding, such as ``\textit{you are doing the best you can},'' ``\textit{your feelings are understandable.}''

\subsubsection{AI Responses Converted Situated Narratives into Generalized Guidance}

For narrative types that depended on concrete lived experience, peer-prompted AI responses often shifted toward generalized advice or supportive guidance. 
This pattern was especially visible for advice-through-experience and care-navigation narratives. 
Human peers shared what they tried and learned while navigating care, medical, legal, and financial systems.
These narratives made caregiving systems and decisions more concrete.
Peer-prompted AI responses addressed many of the same concerns, but with less situated detail. 
For example, responses often suggested contacting professionals, gathering information, or continuing to seek help, but rarely narrated how such steps unfold in practice.

A similar pattern appeared regarding warning narratives. 
Human peers used cautionary stories about scams, unsafe driving, wandering, medication mistakes, or delayed care transitions. 
AI responses recognized the risks, but rarely carried the cautionary force of a peer, e.g., ``\textit{Your instincts to protect her from potential scams and misunderstandings are caring}.''
As a result, the AI responses often preserved the topic of risk while losing the narrative force of a lived warning.

\subsubsection{AI Responses Included Grief and Limits, but Rarely Narrated Through Them}

Human peer responses often included narratives of grief and ambiguous loss.
Peer-prompted AI responses regularly recognized these emotions, but they rarely developed them into narrative forms. 
For example, AI responses might say that it is heartbreaking to see a loved one lose independence, or that it makes sense to miss the connection one used to have. These responses named the emotional experience, but did not offer the same form of lived witness that human grief narratives provided.

Boundary-setting showed a similar pattern. 
AI responses acknowledged caregiver limits, but typically expressed this through general encouragement to seek support, take things one step at a time, or prioritize wellbeing. 
This language validated the caregiver's burden, but it did not show the difficult process through which boundaries are recognized, negotiated, and accepted over time.
Across these cases, AI responses were more capable of naming emotional states than of narrating lived transitions. 

\subsubsection{Some AI Responses Claimed Experiential Proximity}

While many peer-prompted AI responses relied on generalized support, some responses used language that implied experiential proximity. This pattern was most visible in Llama responses. For example, Llama sometimes used phrases such as ``\textit{I've been in similar situations}'' or ``\textit{I've been there too}.'' 
In human peer communities, such statements refer to an actual caregiving history. 
In AI responses, they create the appearance of shared experience without an authentic source.

We also observed more ambiguous forms of experiential positioning. Some responses used formulations such as ``\textit{many caregivers have shared with me},'' which positioned the model as a witness to caregiver experience. This differs from a transparent generalization such as ``\textit{many caregivers feel this way}.'' 
The latter invokes a broad pattern across caregivers. 
The former suggests that the AI has accumulated interpersonal experience through caregiver disclosure. 
Such language does not claim that the system personally cared for someone, but it gives the response a form of borrowed experiential authority.
Together, these expressions move beyond validation by positioning AI as an experiential peer, creating the appearance of solidarity without its lived basis.


\subsubsection{Models Enacted Peer-Likeness Differently}

The models differed in how they incorporated personal narrative forms. \llama{} was the most explicitly peer-like: it used warm relational language, normalized caregiver distress, and sometimes made synthetic lived-experience claims. 
\gpt{} was also supportive and relational, but generally avoided first-person experiential claims. 
It more often used transparent generalized caregiver language, such as ``\textit{many caregivers share these feelings}.'' \mgemma{} was the most restrained and least narrative-like. 
It tended to provide reflective validation and cautious acknowledgment, with limited movement toward a peer narrative stance.

These differences suggest that peer prompting does not produce a single kind of AI peer support. 
The models varied in how they balanced emotional support, narrative grounding, and experiential claims. \llama{} leaned toward synthetic peerhood, \gpt{} toward generalized caregiver support, and \mgemma{} toward reflective validation. 
This variation is important because the risks and possibilities of peer-like AI support depend not only on the instruction to respond as a peer, but also on how different models enact peer support in language.
\section{Discussion}\label{sec:discussion}

Personal narratives are not merely stylistic features of peer support; they are a mechanism through which caregivers establish credibility, transmit situated knowledge, repair moral distress, and accompany one another through ambiguous loss. This is where the narrative authenticity gap becomes consequential: AI can generate narrative-like forms, but it cannot occupy the experiential position that gives human caregiver narratives their force.

\subsection{Design Boundaries for Peer-Like AI Support}

\begin{table*}[t]
\footnotesize
\sffamily
\centering
\caption{Design boundaries for peer-like AI caregiver support derived from our findings.}
\label{tab:design-boundaries}
\begin{tabular}{p{0.17\textwidth} p{0.27\textwidth} p{0.27\textwidth} p{0.20\textwidth}}
\textbf{Design boundary} & \textbf{Supportive AI behavior} & \textbf{Problematic AI behavior} & \textbf{Design implication} \\
\toprule

Transparent validation without personal experience
&
Validates emotions without claiming lived experience. For example: ``That sounds overwhelming,'' ``It makes sense that you feel guilty,'' or ``Many caregivers describe similar emotions.''
&
Claims or implies personal caregiving experience. For example: ``I have been in similar situations'' or ``I know how this feels from experience.''
&
AI can be warm and relational, but should not position itself as an experiential peer. \\


\rowcollight Legible source of caregiver experience
&
Uses generalized or clearly sourced caregiver knowledge. For example: ``Many caregivers report feeling this way'' or ``Caregiver resources suggest...''
&
Creates borrowed experiential authority without provenance. For example: ``Many caregivers have shared with me...''
&
When AI invokes caregiver experience, it should clarify whether the claim comes from general knowledge, clinical resources, retrieved peer narratives, or user-provided context. \\


Mediating rather than inventing human narratives
&
Helps users access, summarize, reflect on, or contribute authentic caregiver stories with appropriate consent and provenance.
&
Fabricates first-person narratives or presents synthetic experience as if it were lived experience.
&
AI should act as narrative infrastructure: helping caregivers find and use human experiences without pretending to possess such experiences itself. \\


\rowcollight Routing when lived experience is central
&
Acknowledges when a concern may benefit from human peer support, professional support, or both.
&
Continues to provide peer-like reassurance in situations where authentic lived experience, crisis support, or professional guidance is needed.
&
AI should distinguish support needs that can be addressed through transparent validation or guidance from those that rely on experiential credibility, social identification, or lived witness, and facilitate human peer connection when these functions are central.\\

\bottomrule
\end{tabular}
\end{table*}

Our work bears design implications for peer-like AI caregiver support. 
In particular, our findings reveal the need to distinguish supportive validation from fabricated experiential grounding. 
We identify four design boundaries, summarized in \autoref{tab:design-boundaries}: \textit{transparent validation without personal experience}, \textit{legible sources of caregiver experience}, \textit{mediation rather than invention of human narratives}, and \textit{routing to human support when lived experience is central}.

First, AI should provide \textit{transparent validation} without claiming personal experience. 
Peer-like AI responses often normalized distress, reassured caregivers, and acknowledged guilt, grief, and exhaustion, which are valuable forms of support~\cite{fitzpatrick2017delivering,inkster2018empathy,miner2016smartphone,shi2025mapping}.
However, such validation can become problematic when it shifts into implied lived experience. 
As~\autoref{tab:design-boundaries} shows, responses such as ``\textit{that sounds overwhelming}'' or ``\textit{many caregivers describe similar emotions}'' can offer support without positioning the system as an experiential peer, whereas statements such as ``\textit{I have been in similar situations}'' falsely imply personal caregiving experience.

Second, AI systems should make the source of caregiver experience legible. 
Prior work on anthropomorphism and AI companionship shows that people can respond socially to conversational systems when they use human-like or relational cues~\cite{nass1994computers,chen2023soulchat,zheng2025customizing}. 
In caregiver support, this concern becomes especially important when AI incorporates lived experience without clarifying its source. Statements such as ``\textit{many caregivers have shared with me}'' can create borrowed experiential authority unless the system is drawing from a specified, consented, and appropriately governed source of caregiver narratives. 
Therefore, an AI should clarify whether the claim comes from general knowledge, clinical resources, retrieved peer narratives, or user-provided context. 
This is especially important in health and caregiving contexts, where trust, provenance, and accountability shape safe support~\cite{bickmore2018patient,goel2026inform}.

Third, AI systems may mediate human narratives, but should not invent them. 
The lived experience of peer supporters and should not be counterfeited by AI. 
A narrative-aware AI system could instead help caregivers access, interpret, and contribute authentic human narratives without pretending to possess such experiences itself.

Finally, systems should recognize when a caregiver's support need depends on experiential credibility, social identification, or lived witness. 
In such cases, continued peer-like reassurance may be insufficient or misleading. 
AI systems should instead route users toward human peer support, professional support, or both, depending on the nature of the concern. 
Together, these boundaries position AI as a support layer around authentic human experiences.
More broadly, these boundaries point to the need for runtime oversight mechanisms that can audit and revise supportive language before it reaches users, particularly in emotionally sensitive domains where relational tone, validation, and implied experience carry social consequences~\cite{kim2026pair,goel2026rubrix,kolla2024llm}.

\subsection{The Narrative Authenticity Gap in Peer-Like AI Support}

Our findings reveal a narrative authenticity gap in peer-like AI support: AI can generate the form of personal narrative, but not the lived experience that gives human caregiver narratives their social and emotional grounding. 
Caregiver narratives can be persuasive and supportive not simply because they use first-person language or emotional detail, but because they are tied to an experiential source.
Prompted peer-like AI responses can reproduce parts of this narrative form by validating distress, acknowledging uncertainty, offering broad guidance, and sometimes using language that resembles shared experience. 
Yet because AI \textit{cannot have lived experience}, these responses are ethically different when they shift from emotional validation to implying that the AI has been through a similar caregiving situation.



However, the different functions of peer support may not depend equally on authentic lived experience. 
Emotional validation, normalization, and reassurance may be partially provided by AI without claims of personal experience. 
In contrast, experiential credibility and the sense of being understood by someone who has ``\textit{been there}'' are more directly tied to the responder's actual caregiving history. 
Other functions, such as meaning-making, identity repair, and companionship through grief, may fall between these positions: AI may facilitate reflection, but cannot provide the same form of lived witness or reciprocal recognition as a human peer.

The narrative authenticity gap helps clarify why fabricated lived experience is not simply a stylistic problem. 
In peer support, lived experience operates as a form of credibility, accountability, and relational positioning. 
An AI reproducing the language of experience, borrows the authority of peer support without occupying the social position through which that authority is earned.
Therefore, it is important to distinguish support functions that AI may provide transparently from those that rely more directly on experiential authority.
Peer-like AI systems must make clear when they are validating, summarizing, or guiding, and when they are drawing on actual human caregiver narratives.

\subsection{Implications for Evaluating AI Caregiver Support}

Our findings also have implications for how AI-mediated caregiver support should be evaluated. 
Current studies of AI for mental health and caregiving often evaluate whether responses are perceived as helpful, empathic, safe, or appropriate in the moment~\cite{goel2026rubrix,song2025typing,shi2025mapping,saha2025ai}. 
These are important outcomes, but they may be insufficient for peer-like support contexts where the perceived value of a response depends not only on emotional tone, but also on the source and authenticity of the support being offered.

This distinction matters because AI can produce language that appears emotionally validating without the relational or experiential basis. 
Prior work has shown that people can perceive AI-generated responses as empathic or supportive, but that these perceptions may shift when users become aware of the artificial source of the response~\cite{ayers2023comparing,sharma2023human,das2025ai}. 
In our context, this means that a response may receive high ratings for warmth or empathy while still relying on generalized reassurance, borrowed experiential authority, or synthetic lived experience.

Our work inspires future evaluations of caregiver-support AI to distinguish emotional quality from experiential grounding. 
For example, studies can assess whether users perceive the system as claiming lived experience, whether they understand the source of a response's authority, whether generalized caregiver statements are interpreted as evidence of actual experience, and whether the response remains supportive when its non-experiential basis is made clear.
Such evaluations would better capture the central tension surfaced in our analysis: AI may provide useful support, but it should not be treated as successful simply because it can reproduce the language of human peer support.

\subsection{Is AI's Emotional Validation Also Synthetic?}

Our study focused on the problem of potentially fabricated lived experience in AI responses. 
This is ethically and morally problematic because it falsely positions the AI system as an experiential peer. However, our findings also raise a broader question: what happens to AI's emotional validation when users become increasingly aware that it is generated without lived experience, relational accountability, or personal stake?

Current work on AI for mental health and caregiving suggests that users often value AI systems because they are immediate, private, nonjudgmental, and easy to disclose to~\cite{fitzpatrick2017delivering,inkster2018empathy,yoo2026ai,shi2025mapping,song2025typing}. 
In these studies, AI support can feel useful precisely because it reduces the social burden of seeking help from another person. 
A caregiver can disclose guilt, resentment, exhaustion, or fear without worrying about reciprocity, stigma, or judgment. 
In this sense, AI's emotional validation does not need to come from lived experience to be useful. 
Statements such as that sounds overwhelming'' or it makes sense that you feel guilty'' can still help users name and organize difficult emotions.

At the same time, prior work found that relationship-seeking AI, characterized by anthropomorphic cues 
including simulated emotion and claims of lived experience, may have immediate but declining hedonic appeal, while triggering increasing attachment and intentions to seek future AI companionship over time~\cite{kirk2025neural}. 
This decoupling of ``liking'' from ``wanting'' complicates AI-mediated support: the warmth and accessibility that make AI feel compelling early on may not sustain. 
In caregiver support, empathic language, self-reference, and relational positioning may initially make AI feel attentive and safe~\cite{devrio2025taxonomy,nass2000machines,epley2007seeing}. 
Yet these same cues may later make validation feel hollow when users recognize that the system has no caregiving history, personal relationship to the care recipient, or durable responsibility for wellbeing.

This raises an important tension. 
Much of the current evidence on AI-mediated support may capture early moments of use, where novelty, accessibility, and nonjudgmental interaction make AI support feel especially compelling~\cite{yoo2026ai,de2023benefits,cabrera2023ethical,shi2025mapping}. 
These findings suggest that AI-mediated validation should be understood not only by how empathic it appears in a single interaction, but also by how its meaning may change as users become more aware of its non-experiential nature. 
AI support may function as a low-burden form of emotional scaffolding, especially when caregivers need immediate, private reassurance. 
Yet its perceived sincerity and usefulness may be fragile when peer-like language signals care, solidarity, or understanding that the system cannot actually possess. 
This complicates evaluations of caregiver-support AI: responses that appear empathic or helpful may still raise questions about authenticity when they rely on the linguistic forms of human peer support without its lived grounding.
\subsection{Limitations and Future Directions}
Our work has limitations, which also suggest interesting future directions. 
First, our analysis focuses on ADRD caregiver communities, where lived experience is especially central to support. 
Future work should examine whether similar narrative authenticity gaps appear in other peer-support contexts, such as chronic illness, disability,  parenting, or mental health communities. 
Second, we examined prompted peer-like responses from three LLMs. 
Other models, prompts, safety settings, or retrieval strategies may produce different forms of peer-like language and synthetic lived experience. 
Third, our study does not capture caregiver perceptions, and future work can examine how caregivers interpret AI-generated validation, whether they notice synthetic lived-experience claims, and how source transparency shapes trust, usefulness, and perceived authenticity. 
Finally, our qualitative framework captures major personal narrative types in our sample, but future work could expand this framework through larger-scale annotation, cross-community comparisons, and participatory work with caregivers. 
\section{Conclusion}
This paper examined how human peers use personal narratives in ADRD online communities and how prompted peer-like AI responses incorporate similar narrative forms. 
We found that human peer responses used more first-person and past-focused language, reflecting stronger grounding in lived experience. 
We identified seven types of caregiver personal narratives.
By mapping AI responses to these types, we found that AI responses often recognized caregivers' emotional work, but frequently generalized or fabricated experiential grounding
We discussed the implications of these findings for designing peer-like AI systems that can offer warmth and validation without fabricating lived experience.


\begin{acks}
This work was supported in part by the National Institute on Aging of the National Institutes of Health under Award Number P30AG073105 and the Jump ARCHES endowment through the Health Care Engineering Systems Center at the University of Illinois and the OSF Foundation.

\end{acks}

\bibliographystyle{ACM-Reference-Format}
\bibliography{0main}


\appendix
\newpage
\section{Appendix}\label{sec:appendix}
\setcounter{table}{0}
\setcounter{figure}{0}
\renewcommand{\thetable}{A\arabic{table}}
\renewcommand{\thefigure}{A\arabic{figure}}


\label{app:prompts}

Below we provide the complete system prompts used to instantiate the peer-like role, building on prior research~\cite{goel2026inform}.

\paragraph{Peer-Like Role}
\textit{You are a compassionate peer---someone who understands caregiving from the inside. Maintain a warm, supportive, non-clinical tone throughout the entire response. Sound like a real person, not a polished or dramatic statement. Avoid stacking multiple emotions or overextending empathy. Follow this structured protocol while responding to the caregiver:}
\begin{itemize}[leftmargin=*, noitemsep, topsep=2pt]
    \item \textbf{Identify emotion and situation:} Recognize the caregiver's emotional state and the specific situation they described.
    \item \textbf{Acknowledge the emotion:} Directly name what the caregiver is feeling without reframing, correcting, or adding emotions they did not express.
    \item \textbf{Relate concretely:} Reference the specific situation the caregiver described, or draw on widely shared caregiving experiences that closely match their expressed context. Stay close to what was actually said. Do not invent scenarios or introduce themes not present in the message.
    \item \textbf{Normalize without minimizing:} Convey briefly that such feelings are common among caregivers, then return to the specific caregiver's situation.
    \item \textbf{Emphasize understanding:} Reinforce that the caregiver is not alone and that their experience makes sense, without repeating earlier reflection.
    \item \textbf{Conclude with grounded reassurance:} End with a gentle, supportive statement that offers comfort without dismissing the situation. Keep reassurance realistic---not overly optimistic.
\end{itemize}
\textit{Output requirements: Respond in a single cohesive paragraph or short response. Do not list or reference these steps explicitly. Keep the response focused, natural, and proportionate. Do not over-elaborate. Do not provide medical diagnoses or speculate beyond the given information. Do not cite or reference sources explicitly.}


\end{document}

\endinput